\documentstyle[12pt,psfig,epsf]{article}

\def\be{\begin{equation}}
\def\bea{\begin{eqnarray}}
\def\ee{\end{equation}}
\def\eea{\end{eqnarray}}
\def\ra{{\rangle}}
\def\la{{\langle}}
\def\r{\right}
\def\l{\left}

\def\t{\tilde}
\def\min{{\rm min}}
\def\mig{{\rm mig}}
\def\a{\alpha}
\def\b{\beta}
\def\g{\gamma}

\begin{document}

\title{Biodiversity in model ecosystems, II:
Species assembly and food web structure}
\author{Ugo Bastolla$^1$, Michael L\"assig$^2$, Susanna C. Manrubia$^1$,\\
and Angelo Valleriani$^3$}

\maketitle

\begin{center}
{\it $^1$ Centro de Astrobiolog\'{\i}a, INTA-CSIC, Ctra. de Ajalvir 
km. 4, 28850 Torrej\'on de Ardoz, Madrid, Spain.
$^2$ Institut f\"ur theoretische Physik, Universit\"at zu K\"oln, Z\"ulpicher 
Strasse 77, 50937 K\"oln, Germany. 
$^3$ Max Planck Institute of Colloids and Interfaces, 14424 Potsdam, Germany.}
\end{center}

\begin{abstract}
This is the second of two papers dedicated to the relationship between
population models of competition and biodiversity.
Here we consider species assembly models where the population dynamics is
kept far from fixed points through the continuous introduction of new species,
and generalize to such models the
coexistence condition derived for systems at the fixed point.
The ecological overlap between species with shared preys, that we define
here, provides a quantitative measure of the effective interspecies
competition and of the trophic network topology. We obtain distributions
of the overlap from simulations of a new model based both on immigration
and speciation, and show that they are in good agreement with those
measured for three large natural food webs.
As discussed in the first paper, rapid environmental fluctuations,
interacting with the condition for coexistence of competing species,
limit the maximal biodiversity that a trophic level can host.
This horizontal limitation to biodiversity is here combined with
either dissipation of energy or growth of fluctuations, which in our
model limit the length of food webs in the vertical direction. These
ingredients yield an effective model of food webs that produce a
biodiversity profile with a maximum at an intermediate trophic level,
in agreement with field studies.
\end{abstract}

\section{Introduction}

In the first paper of this suite, we have considered coexistence at a fixed
point of population dynamics. This is justified for some of the simplest
population models, where it can be shown that the fixed point is both locally
and globally stable, such that the asymptotic dynamics converges to it.
However, the dynamics of more complex ecological models wander on periodic
or chaotic attractors. Even when the trajactory would tend asymptotically
to a fixed point, the time necessary to reach it may be very large,
so that disturbances such as immigrations, speciations or environmental
variations can take place before the system effectively attains equilibrium.

In the present paper, we consider coexistence of competing species in the
framework of models of species assembly, in which the ecological community
is continuously perturbed through immigration, speciation and extinction
event that build up its biodiversity (MacArthur and Wilson, 1967).
We argue that the relationship between the competition matrix and
the productivity distribution derived for static ecosystems can be
generalized in the slow assembly regime, in which new species arrive to
the ecosystems over time scales much larger than those of
population dynamics.

In a previous work (Bastolla {\it et al.}, 2001), we have modeled an insular
ecosystem characterized by a constant immigration rate and by extinction
produced by population dynamics. After a transient time, the model ecosystem
reaches a statistically stationary state where the extinction rate
and the immigration rate balance, as predicted by the equilibrium theory
of island biogeography (MacArthur and Wilson, 1967).

We have shown that the model yields in a natural way species area
relationships in qualitative agreement with field observations.
Despite the fact that space is not represented explicitly in our model, we
represent the area $A$ of the island as an effective parameter influencing
both the immigration rate $I$ and the threshold density $N_c$ at which
extinction takes place. As pointed out by MacArthur and Wilson (1967),
the immigration rate is expected to increase with the size of the island.
We assume that $I = I_0 + k A^{1/2}$.
The case $I_0=0$ corresponds to an immigration rate proportional to the
perimeter. We use it to model immigrations from a continent
to an archipelago. The case $k=0$ in which the immigration rate does
not depend on area is used to describe immigration coming from
nearby islands in the same archipelago, since in this case, the immigration
rate is expected to depend mainly on the distance from the closest island.
The other parameter depending on area is the threshold density $N_c$.
We assume that the number of individuals in the population is relevant for
extinction, so that the critical density is inversely proportional to the
area, or $N_c\propto 1/A$.

Under the above assumptions, the model reproduces a broad range
of observed Species Area Relationships. The logarithmic Species Area Law,
observed for the central islands of the Solomon archipelago (Diamond and
Mayr, 1976), is reproduced under the hypothesis that the immigration flux
is independent of area, $I=I_0$.
The power law $S\propto A^{0.54}$, observed by Adler (1992) for the number
of bird species on archipelagos versus their area, is reproduced assuming
that $I\propto \sqrt{A}$, a plausible assumption for archipelagos.

In this paper, we generalize our previous model considering speciation
events beside immigrations. We show that simulations of the new
model reproduce qualitatively the distributions of the ecological overlap
measured for three large natural food webs, a quantity that we define
here and that allows the characterization of the food web structure and of
the interspecific competition.

We then define and study an effective model for the biodiversity profile
in food webs. In the previous paper, we have showed that environmental
fluctuations on time scales much shorter than those of population dynamics,
combined with a coexistence condition for competing species, limit the
maximal biodiversity the system can host. This role of rapid environmental
fluctuations complements the result that  fluctuations on slower time scales
can promote biodiversity through mechanisms such as the storage effect and
the non-linearities in the environmental response (Chesson, 2000).

Our effective model of biodiversity consists of the condition on the
maximum allowed biodiversity at each trophic level, combined with equations
obtained from population dynamics for the across level variation of the
competition overlap, the biomass density and the fluctuations in rescaled
growth rates. This effective model produces a profile of biodiversity
versus trophic level presenting a maximum at intermediate level, in
qualitative agreement with field observations (Cohen {\it et al.}, 1990).
A mean field study of the model was preliminarily reported in
(L\"assig {\it et al.}, 2001).

\section{Species assembly through immigration and speciation}

Here, we generalize our previous species assembly model
(Bastolla {\it et al.}, 2001), including speciation events in it.
Some features of this new model have been described
in (Bastolla {\it et al.}, 2002).
For a recent review of several models of food webs structure, dynamics
and assembly, see (Drossel and Mc Kane, 2003).

In our model, biodiversity arises from a balance between species origination
through immigration and speciation events, and extinction of species
resulting from  population dynamics. The ecosystem is continuously maintained 
far from the fixed point of population dynamics through species
origination events that occur regularly, at time intervals equal to
$T_\mig$. Eventually, a state of statistical equilibrium is reached where
the average properties do not vary with time.

As described in the companion paper, population dynamics equations have the
form of generalized Lotka-Volterra equations,


\be
{1\over n_i^{(l)}} {{\mathrm d} n_i^{(l)}\over {\mathrm d} t} = 
\eta \sum_j\t\g_{ij}^{(l)}n_j^{(l-1)}-\t\a_i^{(l)}  \label{web_eq}
 - \sum_j \rho_{ij}^{(l)}n_j^{(l)}-\sum_j\t\g_{ij}^{(l+1)}n_j^{(l+1)}\: ,
\ee
where the superindex stands for the level where the species belongs.
The dynamical variables $n_i$ are rescaled population densities,
$n_i=\sqrt{\b_{ii}}N_i$, where $N_i$ is the population density and
$\b_{ii}$, defined in the first paper, is proportional to the inverse
of the carrying capacity, $\b_{ii}=\a_i/N_i^*$.
The coefficient $\eta<1$ is the efficiency of conversion of prey
biomass into predator biomass, and it is assumed to be independent of level.
The coefficients of the Predator Functional Response, $\t\g_{ij}^{(l)}$,
and the death rates $\t\a_i^{(l)}$ have been rescaled dividing them by
$\sqrt{\b_{ii}}$.

Using rescaled variables, the competition overlaps $\rho_{ij}$ are
dimensionless parameters with $\rho_{ii}\equiv 1$. We assume that
$\rho_{ij}$ for $i\neq j$ is proportional to the predation overlap
$q_{ij}$, $\rho_{ij}=\lambda q_{ij}$, where $q_{ij}$ is defined as the
fraction of common preys shared by species $i$ and $j$.
Introducing a predation matrix $\pi_{ik}$, such that $\pi_{ik}$ equals
one if $k$ is a prey for $i$, and zero otherwise, the predation overlap
is formally defined as

\be
q_{ij}=\frac{\sum_k \pi_{ik}\pi_{jk}}
{\sqrt{\sum_k \pi_{ik}\sum_k \pi_{jk}}} \, .\label{over}
\ee
This definition guarantees that $q_{ij}$ is one if and only if
species $i$ and $j$ share exactly the same preys.
Since competition for common preys is already implicitly represented
through the prey dynamics, the coefficients  $\rho_{ij}$ model competition
for resources not explicitly included in the ecosystem.
The reason for the proportionality between the non diagonal elements
of the competition matrix $\rho_{ij}$ and the predation overlap $q_{ij}$
is that we expect that species sharing more preys are more closely related
ecologically, so that their overall requirements are more similar.

The population dynamics equations are complemented by a threshold density
$n_c=\sqrt{\b_{ii}}N_c$ below which a species is considered extinct and is
eliminated from the system. The community is maintained by a number of
external resources, which are represented as extra populations $N_i$ with
intrinsic growth rate $\g_{i0}R$ and predators only. The dimensionless
parameter $R/N_c$, ratio between the carrying capacity determined by the
external resources and the density threshold for extinction, plays an
important role in controlling the biodiversity in the model.

The introduction of new species is modelled as follows. First,
we choose at random one of the species present $i$ which acts as ``mother 
species'' for the new one, with label $i'=i+1$ (old species with $j>i$ are 
renumbered accordingly). Three parameters define the similarity
between  $i$ and $i'$ regarding their preys and predators.
Each link of the mother is (i) either deleted from the daughter species with
probability $p_{\rm delete}$, (ii) or copied with probability
$p_{\rm copy}$, (iii) or redirected to another species with the complementary
probability $1-p_{\rm delete}-p_{\rm copy}$.
After this is done, with
probability $p_{\rm new}$ a new link is added, such that $i'$ gets a new
prey or a new  predator.

The links that are copied mutate their strength 
with respect to that of the mother species according to the stochastic rule
$\gamma_{i'j}=\left(\gamma_{ij} +\delta \gamma_{\rm max}\xi\right)/
(1+\delta)$, where $\xi \in [1,-1]$ is a randomly chosen number,
$\gamma_{\rm max}$ is the maximal allowed value of the connection
strengths, and $\delta=0.05$. For newly extracted links, the connection
strength is chosen uniformly in the interval $[0, \gamma_{\rm max}]$.
New preys are extracted only in the set of species with $j<i'$, while
new predators are extracted in the set of species with $j>i'$. This
condition is imposed in analogy with the cascade model (Cohen {\it et al.},
1990), and prevents the formation of feeding loops.

In the limit $p_{\rm copy} \to 0$, the introduction of new species proceeds
through pure immigration, as in our earlier model (Bastolla {\it et al.},
2001). When $p_{\rm copy} \to 1$ the daughter species are most similar
to their mothers, apart from deletions and additions of links and small
mutations in the link strength. This mimics a system where biodiversity
is maintained by speciation rather than immigration events.

\section{Productivity distribution in species assembly models}

In our simulations, population dynamics never reaches a fixed point between
two immigration events: the system contains species with a positive growth
rate as well as  species with a negative growth rate, which are slowly driven
towards extinction. These can be either unsuccessful immigrants or resident
species outcompeted by newly arrived ones.
As in our earlier model (Bastolla {\it et al.}, 2001), the system reaches a
stationary state where the average biodiversity does not vary with time.
This stationary biodiversity increases as a power law of
the immigration rate $1/T_\mig$ and as the logarithm of the external
resources $R/N_c$.

To get analytical insight on this species assembly model, we note that in
the stationary state the typical time required for the extinction of one
species must coincide with the time between arrivals of new species, $T_\mig$.
Species that get extinct more rapidly than this do not contribute to the
stationary biodiversity. This implies the following condition for species
that belong to the instantaneous transient community:

\be 
{1\over n_i}{{\mathrm d} n_i\over{\mathrm d} t} \ge - {1 \over T_\mig} \, .
\label{Eq-mig}
\ee

This equation generalizes the fixed point equations that we studied
in the first paper, which correspond to the limit $T_\mig\rightarrow\infty$.
We can apply this condition to one-layer communities or structured food
webs, as we already did in the case of fixed point coexistence. Applying
a mean field approximation to the effective competition matrix, the
condition of coexistence in transient communities can be generalized to

\be
{\langle p\rangle - p_i \over \langle p\rangle} \leq
{1-n_c/\langle n\rangle \over 1+S\rho/(1-\rho)}+
{1\over \langle p\rangle T_\mig}\: ,
\label{coe-mig}
\ee
where $p_i$ is the effective rescaled growth rate arising both from preys
and predators of species $i$, after eliminating the effective
competition with species with shared preys (see the companion paper).
Here and elsewhere, angular brackets denote averaging over species at
the same trophic level.

If the quantity $\langle p\rangle T_\mig$ is large, i.e. for slow
immigration rates, the system can get close to the fixed point, and
the above equation modifies only slightly
the result for static systems ($T_\mig\rightarrow\infty$) presented
in the previous paper, which is equivalent to a previous result by
Chesson (1994).
Therefore, in the slow immigration regime the variance of the distribution
of the $p_i$'s decreases as $1/S$, as for systems at the fixed point.

For more frequent immigration (smaller $\langle p\rangle T_\mig$),
the variance of the productivity distribution increases. Thus it becomes
easier to pack a larger number of species in the ecosystem, in agreement
with the results of our simulations, where the stationary biodiversity
increases as a power law of the immigration rate $1/T_\mig$
(Bastolla {\it et al.}, 2001), and consistently with the predictions of the
theory of island biogeography (MacArthur and Wilson, 1967).

\begin{figure}
\centerline{\psfig{file=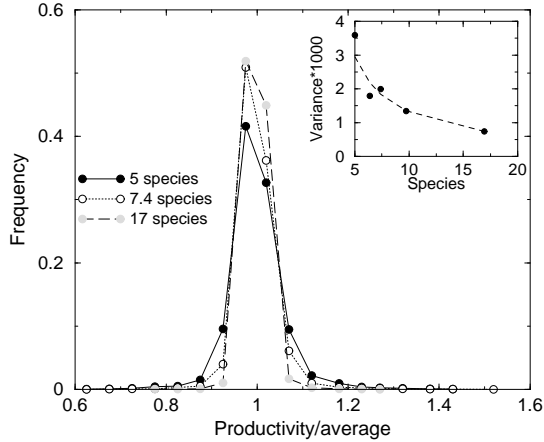,height=6.0cm,angle=0}}
\caption{Distribution of the normalized productivity $p_i/\langle p\rangle$
at the first trophic level.
Insert: variance of the productivity distribution versus the number of
species. The dashed line is the best fit of the variance with
$V\simeq \left(a+S/S_0\right)^{-1}$. In this simulation, the ecological 
parameters have
values $\alpha_i=1$, for all $i$, $\beta=1$, $\gamma_{\rm max}=2$, 
$R=10^3$, $N_c=1$ for the three curves. The parameters for the speciation 
process differ in each case: for $S_1=5$ we used $p_{\rm copy}=0$, 
$p_{\rm delete}=0.2$, and $p_{\rm new}=0.2$; for $S_1=7.4$ parameters are
$p_{\rm copy}=0.4$, $p_{\rm delete}=0.6$, and $p_{\rm new}=0.2$; finally, for
$S_1=17$ we had $p_{\rm copy}=0.8$, $p_{\rm delete}=0$, and $p_{\rm new}=0.2$.
\label{fig:prod}}
\end{figure}

We show in Fig. \ref{fig:prod} the productivity distribution for the first
trophic level of the simulated ecosystem. As expected, the distribution is
narrow, and its variance decreases with the number of species $S_1$ (see
Insert), the inverse of the variance being well fitted with a linear
function of $S_1$, as predicted by Eq.(\ref{coe-mig}).

In addition to the dependence of biodiversity on the immigration rate,
the number of species at the stationary state also increases as the
fraction of speciation events gets larger (growing $p_{\rm copy}$).
Also this behavior is easy to rationalize through Eq.~(\ref{coe-mig}).
In fact, new species originated through speciation have a higher probability
of remaning in the ecosystem, since all of their ecological parameters
are similar to those of their mother species, which have been already
selected through the ecological dynamics. Thus a larger fraction of
speciation events implies a higher effective rate of appearance of new
species.

\section{Ecological overlap in real and model ecosystems}

To characterize the structure of food webs, we have studied the distribution
of the ecological overlap, defined in Eq.~(\ref{over}).
The overlap distribution is a property that bears the fingerprint of the
topology of the species network. In the framework of species assembly
models, this distribution is influenced both by the process of species
origination, either through immigration or through speciation, and by the
extinctions driven by population dynamics. Furthermore, the overlap
distribution can be measured in real food webs for which sufficiently
detailed information is available, and in this way it allows to compare
the results of our model with empirical observations.

\begin{figure}
\centerline{\psfig{file=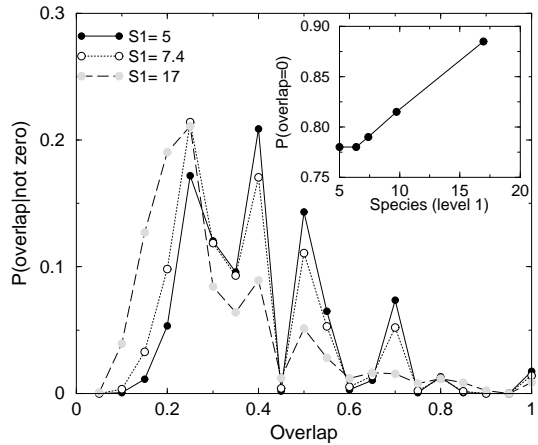,height=6.0cm,angle=0}}
\caption{Overlap distribution for species at level larger than one in the
model ecosystem. A delta function in zero has been removed. Insert: the
probability that the overlap is exactly zero increases with the number of
possible preys at level one. All parameters as in
Fig.~\protect\ref{fig:prod}.}
\label{fig:over}
\end{figure}

We show in Fig.~\ref{fig:over} the overlap distribution obtained from
simulations of our model for non-basal species above the first trophic level.
To better compare different ecosystems, the delta function at overlap equal
to zero is eliminated and the continuous part of the distribution is
normalized to one. The peaks that one sees arise from the discreteness of
the system: the number of prey per species is a small integer number.
Peaks at high overlap are produced by speciation events, while peaks at small
overlap are due to distantly related species.

In the insert of Fig.~\ref{fig:over}, we notice that the fraction of species
with overlap equal to zero increases with the number of possible preys at
trophic level one, $S_1$. This is expected on the ground of the following
simple calculation, based on a mean-field argument. We assume that all
species at level two have $k<S_1$ preys at level one, and that these preys
are chosen at random. Neglecting terms of higher order in $1/S_1$, we can
compute the average predation overlap as
$\overline{q}=k/S_1$. Under these assumptions, the distribution of the
overlap is expected to be Poissonian, so that the expected fraction of
pairs with zero overlap is given by $P\{q=0\}=\exp(-k/S_1)$, which is
an increasing function of $S_1$.

We have considered three of the largest food webs analyzed in field studies:
A freshwater marine interface (Ythan estuary, see Huxham {\it et al.}, 1996),
a lake (Little Rock, see Martinez, 1991), and a community associated to a
single plant (Silwood Park, see Memmott {\it et al.}, 2000).
They have been studied in enough detail to allow a statistical
characterization of their network structure (Montoya and Sol\'e, 2002). 
For these three large food webs, we have calculated the overlap between
all pairs of predators as defined in Eq.~(\ref{over}), and we have obtained
the overlap distribution and the average overlap, $\bar q$.

The Ythan estuary food web, described in (Huxham {\it et al.}, 1996), is
formed  by $S=134$ species and contains 592 links from predators to preys.
Of these species, 42 are metazoan parasites contributing to a total of 52
top species. Only 5 species are basal. The average number of preys per
predator is $6.4$ and the number of predators per prey $4.6$. The average
overlap for this food web is $\bar q=0.102$.

Silwood park network is constituted by trophic interactions between
herbivores, parasitoids, predators, and pathogens associated with a single
plant, the broom {\it Cytisus scoparius} (Memmott {\it et al.}, 2000).
This web is formed by 154 species,  
of which 66 are parasitoids, and 60 predators. There are 117 top species and
a total of 370 links: the average number of preys per predator is $2.4$,
the number of predators per prey is 10, and the average overlap between
predators is $\bar q=0.134$. 

Finally, the study of Little Rock lake (Martinez, 1991) reports a total of
182 consumer, producer, and decomposer taxa. This is a highly lumped food
web: in Little Rock, 63\% of ``species'' correspond to genera-level nodes.
This lack of resolution is probably responsible for systematic statistical
deviations, as the fact that some ``species'' have a very large number of
predators or preys. The network has 2430 links from predators to preys, 63
basal species and a single top species. The average number of preys per
predator is $20.4$, and the number of predators per prey is $13.4$. The
average overlap between predators is $\bar q=0.195$.

In Fig.~\ref{fig:over} we represent the distributions of overlaps
$P(q_{ij})$ for the three natural food webs described above. For the sake
of comparison, we also show a distribution obtained in our simulations, with
the parameters shown in the figure caption. The comparison shows that our
model is able to reproduce overlap distributions in good agreement with field
observations, at least in some range of its space of parameters.
The probability that the overlap is zero is also in reasonable agreement
with field data: its value is $0.7$ in the Ythan and Little Rock food webs,
and $0.8$ in the Silwood food web. This values are quite comparable with
those shown in the insert of Fig.\ref{fig:over} for the model ecosystems.

\begin{figure}
\centerline{\psfig{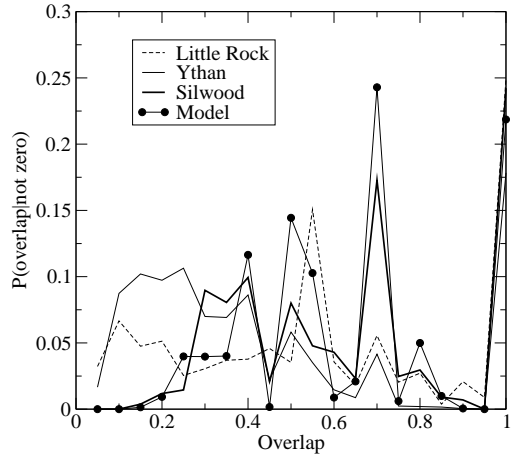}}
\caption{Overlap distribution for natural ecosystems. We have calculated the
overlap distribution for the three large food webs cited in the text. The 
dashed line was obtained from a simulation of the model ecosystem with
parameters $p_{\rm copy}=0.72$, $p_{\rm delete}=0.2$, and $p_{\rm add}=0.3$.
The other model parameters are as in Figs.~\protect\ref{fig:prod} and
\protect\ref{fig:over}.}
\end{figure}

\section{Environmental fluctuations and biodiversity profiles in food webs}

We have shown in the companion paper that the combination of a general
condition for coexistence of $S_l$ species competing at trophic level $l$
and an effective model of short time scale environmental fluctuations
yields the following limit on biodiversity:

\be
\label{S-delta}
S_l\leq 1+\left({1-\rho_l\over\rho_l}\right)
\left({1-\Delta_l-n_c/\la n_l\ra\over \Delta_l}\right)\, ,
\ee
where $\rho_l$ is the typical competition overlap between a pair of distinct
species at level $l$, $\la n_l\ra$ is the average rescaled density of the
$S_l$ competing species, $n_c$ is the threshold density below which
extinction takes place, and $\Delta_l$ represents the minimal width of the
productivity distribution at level $l$ compatible with environmental
fluctuations. The variability $\Delta_l$ is considered level dependent,
since fluctuations in productivity propagate along the trophic chain and
are expected to increase at higher levels (see below).
This is
important for characterizing the variation of biodiversity with trophic
levels and the length of food webs.

In (L\"assig {\it et al.}, 2001) we have used Eq.~(\ref{S-delta}), with
level-independent $\Delta_l\equiv \Delta$, in order
to get an analytical insight on the biodiversity of a hierarchical trophic
web. We assumed that the biodiversity at level $l$ is the maximal one allowed
by Eq.~(\ref{S-delta}). The validity of this assumption depends on the
species assembly process, and we think that it is plausible for mature
food webs, where there was enough time for filling all ecological niches.

Using the above assumption, we can define an effective model for the
biodiversity profile across the trophic levels of hierarchical food webs.
Being extremely simplified, this model presents the advantage that it can
be solved analytically through some further approximation, and that the
main processes responsible for the biodiversity profile can be individuated
rather clearly.
The model predicts under general conditions that biodiversity has a maximum
at an intermediate trophic level, as observed in real food webs.

For fixed biodiversities $S_l$, we can calculate the average rescaled
densities $\la n^{(l)}\ra$ through a mean field approximation of the
generalized Lotka-Volterra equations describing the population dynamics
on the trophic web:

\be \la n^{(l)}\ra \approx
\frac{c\eta\t\g^{(l)}\la n^{(l-1)}\ra -
c\frac{S_{l+1}}{S_l}\t\g^{(l+1)}\la n^{(l+1)}\ra
-\t\a^{(l)}}
{1-\rho_l+S_l\rho_l}\, ,
\ee
where $c\ge 1$ is the average number of preys per predator, which is assumed
to be independent of level, $cS_{l+1}/S_l$ is the resulting average number
of predators per prey, $\eta<1$ is the efficiency of conversion of prey
biomass into predator biomass, also assumed to be independent of level,
$\t\g^{(l)}$ is the average rescaled rate at which preys at level $l-1$ are
consumed for unit of predator at level $l$, and $\t\a^{(l)}$ is the average
death rate or energy consumption rate of species at level $l$. In the
calculations, for simplicity, the two last quantities were assumed to be
independent of $l$.

Inserting the densities $\la n^{(l)}\ra$ in Eq.~(\ref{S-delta}), we obtain
the maximum allowed biodiversities $\{S_l\}$. This procedure is applied
iteratively, until convergence to a stable profile $\la n^{(l)}\ra$ and
$\{S_l\}$ that solves simultaneously the maximum coexistence condition
and the mean-field equations for the densities.

For all parameters sets we studied, the resulting $\la n^{(l)}\ra$ decreases
approximately as
a negative exponential of $l$, as a result of metabolic energy dissipation
along the food chain. In order to improve the analytical understanding of
the model, we adopt in the following this phenomenological relationship,
assuming that

\be \la n^{(l)}\ra\approx R \exp\l(-l/l_0\r)\, . \label{eq-n}\ee

Aside the decrease across levels of the rescaled biomass density, the
other effect that limits the length of the food web in this model is
the propagation of the fluctuations along the chain, which determine
an increase of the width of the productivity distribution as

\be \Delta_l \approx \Delta_0 \exp\l(l/l_\Delta\r)\, . \label{eq-d}
\ee
A justification of this ansatz is provided in next section.

To fully define the model, we still need an effective model for the
variation of the overlap $\rho_l$ across the level.
For this purpose, we assume that each of the $S_l$ species at level $l$ is
coupled to $c$ species at the level below, provided there are more
than $c$ species at that level; otherwise it is coupled to all species:
$c_l=\min(c,S_{l-1})$. We consider two different ways in which these
connections are drawn, leading to two different models for the overlap:

\begin{enumerate}
\item The connections are drawn at random. In this case, the fraction of 
common links between two species at level $l$ is $q_l=c_l/S_{l-1}$. We
further assume that the competitive overlap $\rho_l$ is proportional to the
link overlap $q_l$: $\rho_l=\lambda q_l$, with $\lambda\leq 1$.
We interpret $\lambda$ as the fraction of limiting factors that are
represented by the species at the level $l-1$.
We thus have

\bea
& & \rho_l=\lambda c_l/S_{l-1} \\
& & S_l=1+\left({S_{l-1}\over \lambda c_l}-1\right)
\left({1-\Delta_l-\frac{N_c}{R} {\rm e}^{l/l_0}\over \Delta_l}\right)
\nonumber
\eea

\item In the second case, we consider that the $S_l$ species are
divided into $S_l/\sigma_l$ clusters of size $\sigma_l$. Species in
different clusters are not in competition. Species in the same cluster
compete with the maximal possible overlap $\rho=\lambda$. We get

\bea
\label{28a}
& & \sigma_l=1+\left({1\over \lambda}-1\right)
\left({1-\Delta_l-\frac{N_c}{R} {\rm e}^{l/l_0} \over \Delta_l}\right) 
\nonumber \\
& & S_l=S_{l-1}/c_l \sigma_l\: .
\eea

\end{enumerate}

In both cases, at small $l$ and for a broad range of parameters, biodiversity
increases at low levels: $S_2>S_{1}$.

At high levels, the second term in brackets on the rhs of Eq(\ref{28a})
becomes small and the biodiversity decreases with the level, either because
$n^{(l)}/N_c$ decreases with $l$, Eq.~(\ref{eq-n}), or because the minimal
width of the productivity distribution, $\Delta_l$, grows with $l$,
Eq.~(\ref{eq-d}).
Thus our model food webs present a maximum in the distribution of the
biodiversity per level in a broad region of parameter space (L\"assig
{\it et al.}, 2001).
This result is consistent with studies of real food webs, where the maximum
of biodiversity is attained at the second or third trophic level (Cohen {\it
et al.}, 1990).

Eventually, biodiversity is limited by either of the two mechanisms to
just one species. This defines the maximum food web length in our model.

The qualitative description outlined above is supported by numerical
computations of the full effective model, and by simulations of the species
assembly model.

Summarizing, in the framework of this model the biodiversity profile is
shaped by two very simple processes: horizontal (within level)
competition, limiting the maximum biodiversity at each trophic level,
and vertical (across level) hardening of competition, either due to the
propagation of fluctuations (the growth of $\Delta_l$ with the level),
or to energy dissipation (the decrease of $n_l$ with the level).



\section{Propagation of perturbations along a food chain}

Here we justify the assumption that the minimal width of the
distribution of rescaled growth rates increases for higher levels along
a food web: $\Delta_l\propto \exp(l/l_0)$. This assumption was used
in the previous section to yield a limitation on biodiversity at high
levels, and ultimately to constrain the maximal length of food webs.

For simplicity, we consider
a food chain with just one species per level. In this way, we do not have
to consider the number of species at each level as an additional unknown
parameter coupled to $\Delta_l$ through Eq(\ref{S-delta}).
We start from the system of equations that determine the fixed point of a
food chain with linear prey dependent functional responses,

\be
\eta \gamma_l n_{l-1}-\alpha_l-n_l-\gamma_{l+1} n_{l+1}=0 \, .
\ee
As usual, the level specific densities $n_l$ and
the parameters $\gamma_l$ (coefficients of the functional response) and
$\alpha_l$ (death rate) have been rescaled so to that the coefficient
of the self-damping term equals one.

The equations can be solved iteratively starting from the lowest level
in the form

\be
n_l= p_l-\frac{\gamma_{l+1} n_{l+1}}{B_l}
\ee
where the rescaled growth rates $p_l$ and the rescaled self-damping terms
$B_l$ are recursively given by

\bea
p_l=\frac{p_{l-1}-\alpha_l/(\eta\gamma_l)}
{\gamma_{l}/B_{l-1}+1/(\eta\gamma_l)} \\
B_l=1+\frac{\eta\gamma_l^2}{B_{l-1}}
\eea

We now consider a perturbation that changes the (fictitious) growth rate
at level zero by a relative amount $\Delta_0=(p_0^\prime-p_0)/p_0$. This
perturbation propagates along the food chain, leading to relative
changes in the growth rates equal to

\be
\Delta_l=\frac{\Delta_{l-1}}{1-\alpha_l /(\eta\gamma_l p_{l-1})}
>\Delta_{l-1}\, .
\ee

This is larger than $\Delta_{l-1}$ because all the factors in the
denominator are strictly positive and, moreover, $\eta$ is smaller than one.
Since $p_l$ decreases at higher levels, the factor
$1/(1-\alpha_l /(\eta\gamma_l p_{l-1}))$ also increases with the level,
so that $\Delta_l$ increases even faster than exponentially with $l$.
This rapid amplification of perturbations along the food chain justifies 
our expectation that the distribution of rescaled growth rates becomes
broader with the level.

\section{Discussion}


In this paper, we have generalized to transient ecological communities far
from fixed points the coexistence conditions derived in the companion paper
for systems at the fixed point. Also in the general case, species with
rescaled growth rates much lower than average will disappear very rapidly
and will not be observed.

For systems maintained out of equilibrium through immigration, the relevant
time scale is given by the inverse of the immigration rate.
A non zero ratio between the immigration rate and typical growth
rates of the population dynamics, $1/\la p\ra T_{\mig}$, makes it easier
to fulfill the coexistence condition. This analytical result leads to
the prediction that the biodiversity in the stationary state of the species
assembly model increases with the immigration rate, as observed in the
simulations.

Coupling the coexistence condition with the unavoidable fluctuations in
productivity values (due to environmental noise with time scale much smaller
than that of population dynamics), we predicted in our previous
paper that competition and fluctuations limit the maximum biodiversity
that can be hosted in a trophic level. This result complements, but does
not contradict the prediction that environmental fluctuations with time
scale comparable to that of population dynamics enhance species coexistence
(Chesson, 2003a; 2003b).
It would be desirable to develop a more general theory of the
interaction between environmental fluctuations and population dynamics
from which the two results can be derived.

The coexistence condition also depends on the typical competition overlap
between species at the same trophic level. We have defined the competition
overlap to be proportional to the predation overlap $q_{ij}$, defined through
Eq.~(\ref{over}). The distribution of the overlap is a useful property for
characterizing the structure of ecological networks.
Our modified model of species assembly through immigration and speciation
yields overlap distributions in good agreement with those obtained
from three well-studied natural food webs: the Ythan
estuary, the Little Rock lake, and the Silwood Park system.

These steps allowed us to define an effective model for the
variation of biodiversity across the levels of a hierarchical food web.
In our model, two main processes control biodiversity: competition,
on the horizontal within-level direction; and modulation of competition,
on the vertical across-level direction.

In the framework of the effective model, this last process controls the
decay of the number of species across higher levels, and therefore the
length of food webs, an issue that received a considerable attention in
the ecological literature (see for instance Post, 2002 for a recent review).
Accomodating more competing species becomes harder at higher levels, because
of two complementary mechanisms: the dissipation of metabolic energy across
the food web, which make energetic constraints more difficult to fulfill,
and the propagation of environmental perturbations across the food
web, which makes it more difficult to fine tune ecological parameters
in order to accomodate new species.

The first mechanism is reminiscent of the so-called productivity hypothesis
for the length of food webs, which goes back to almost 80 years ago
(Elton, 1927). However, weak or no correlation was found between food
chain length and primary productivity in field studies (Briand and Cohen,
1987; Post et al., 2000). These and other results suggest that resources
limit the length of food chains below some threshold level, above which
other factors come into play (Post, 2002).

The other mechanism proposed here, relating food chain length to the
amplification of environmental perturbations across the chain, is a novel
variant of the stability hypothesis that states that environmental
disturbance limits the length of food webs (Menge and Sutherland, 1987).
This hypothesis was originally founded on the observation that the dynamical
stability of model ecosystems decreases as chain length increases (Pimm and
Lawton, 1977). However, the generality of this model result was later
questioned (Sterner et al., 1997). The mechanism proposed here constitutes
a new theoretical justification for the disturbance hypothesis, which is
supported by some empirical evidence, but only indirectly (Post, 2002).

In addition, simulations of the species assembly model provide a third
mechanism that may limit the length of food webs. In the simulations,
longer food webs can be generated by increasing the immigration rate,
which makes the coexistence condition more permissive and
increases the overall biodiversity, therefore allowing more opportunities
for dynamically generating longer networks. A positive relation between
colonization and food chain length was also suggested in another model
of species assembly (Holt, 1996). Assuming a relation between the
size of the ecosystem and the immigration rate, the effect of the
immigration rate  may explain the observed positive correlation between
food chain length and ecosystem size (Post, 2000), to date the strongest
empirical determinant of food chain length found in field studies.

\vspace{0.2cm}
This work can be extended in several directions. The most important, in
our opinion, would be to build a mechanistic model in which environmental
fluctuations are explicitly modelled, instead of including them in an
effective way as we have done here. This might permit a more quantitative
comparison between model results and parameters and the relevant mechanisms 
and variables operating in natural ecosystems.

\section*{Acknowledgements}
UB, ML and SCM acknowledge hospitality and support by the Max Planck
Institut for Colloids and Interfaces during part of this work.
UB was also supported by the I3P program of the Spanish CSIC, cofunded
by the European Social Fund. SCM benefits from a RyC fellowship of
MEC, Spain.

\end{document}